# Two Power Series Models of Self-Similarity in Social Networks

Subhash Kak

**Abstract**
Two power series models are proposed to represent self-similarity and they are compared to the Zipf and Benford distributions. Since evolution of a social network is associated with replicating self-similarity at many levels, the nature of interconnections can serve as a measure of the optimality of its organization. In contrast with the Zipf distribution where the middle term is the harmonic mean of the adjoining terms, our distribution considers the middle term to be the geometric mean. In one of the power series models, the scaling factor at one level is shown to be the golden ratio. A model for evolution of networks by oscillations between two different self-similarity measures is described.

## 1. Introduction

A social network consists of N nodes, labeled 1, 2, … n, that may be people, firms, or other entities. The network is represented by a graph where the connection between the nodes *i* and *j* is shown by a link between the two. A network may be valued for the function it performs to nodes outside of it, or it may be valued for the function it serves to the nodes within (Aral and Walker, 2012). Unlike engineered networks whose function (to nodes outside the network) is well-defined, goodness of social networks cannot be easily quantified because the entity that is scarce is attention which requires a different kind of economics (Simon, 1971; Goldharber, 1997; Essig and Arnond, 2001; Yu and Kak, 2014) and also because of the elusiveness of the inner experience of well-being. Nevertheless, for the entire network one speaks of some general function such as value, utility, well-being, or welfare of the group whose definition is driven by extraneous considerations of theory or ideology (Kirman, 1997).

In economic networks (Bienenstock and Bonacich, 1993; Ellison, 1993; Slikker and van den Nouweland, 2000) where value must be associated with the nodes within, one speaks of Pareto efficiency, or Pareto optimality, which is a state of allocation of resources in which it is impossible to make any one individual better off (in terms of a suitable measure of well –being) without making at least one individual worse off. A state is Pareto efficient or optimal when no further improvements can be made. The allocation, normally in terms of resources, could also be in terms of some property of the connectivity. We are not concerned with the cost of the connections and interested primarily in natural connectivity as determined by fundamental considerations.

If one were considering connectivity as value, the total value of the network is proportional to *n* (*n* −1), that is, roughly, $n^2$ since in a network with n nodes each can make (n −1) connections with other nodes. This ignores the fact that in an evolutionary network the connection capacity of the nodes must vary as new nodes can only gradually get connected with others and some other nodes might leave the network. A careful analysis indicates that the potential value of a network of size *n* grows in proportion to *n* log *n* which is connected to the need for self-similarity in natural systems (Briscoe et al., 2006).

Any theoretical definition of utility for a social network is at best arbitrary; therefore the actual connectivity can provide insight into value. Such connectivity is determined by cognitive, sociological, cultural and economic considerations, but one defining feature of it appears to be self-similarity across different layers. It is significant that self-similarity shows up in a variety of complex systems such as the World Wide Web, social, biological, cellular and protein-protein interaction networks and these are well characterized by power-law distributions. The logic behind the emergence of self-similarity in such networks is that of local and hierarchical interactions. The value of the network may be examined from how close it is to self-similarity across many different layers. This may be contrasted with theories that argue that value can be freely transferred across individuals in the society and, therefore, a sign of an efficient network is the total of the values of the individuals. Since value is relative and enhanced by scarcity, aggregate wealth cannot be a good measure of the success of a society.

This paper presents an approach to investigating structure in social networks from the perspective of self-similarity, which one may assume leads to social efficiency. Two power series models have been used for this idea and they have been compared to the Zipf and Benford distributions. In contrast with the Zipf distribution where the middle term is the harmonic mean of the adjacent terms, our distribution considers the middle term to be the geometric mean. A model for the evolution of networks by means of oscillation between two modes is advanced. This may be contrasted with the preferential attachment model of generating scale-free networks. An internal mechanism involving two modes is desirable because it represents an inner tension that drives the dynamics.

## 2. Scale-free and small-world properties

Complex physical and biological networks are characterized by the small-world and the scale-free properties (Albert et al., 1999; Dutta et al. 1998; Watts, 2001; Watts and Strogatz, 1998). In a small-world network one can reach a given node from another one in a small number of steps that are characteristic of social networks. For a complex network, the probability distribution of the number of links per node, *p(k)*, is generally given by a power-law (scale-free) with a degree exponent $\lambda$ usually in the range $2 \leq \lambda \leq 3$:

$$p_1(k) = ck^{-\lambda}, \quad k = m, m+1, ..., K \tag{1}$$

where *c* is for the normalization of the expression (1) and *m* and *K* are the lower and the upper cutoffs of the distribution. The value *k=K* represents the nodes with the largest number of connections which have the least probability. As *k* becomes large, $p_1(k) \approx \sqrt{p_1(k-1)p_1(k+1)}$, and this points to the geometric nature of this distribution for such values. Power-law distributions have long tails.

In the related Zipf's law for natural languages (Zipf, 1949), the frequency of a word is inversely proportional to its rank in the frequency table: $p_z(k) \propto k^{-1}$ (*k* is the rank; for English the most commonly occurring word is "the" which as a probability of about 7% in the Brown Corpus). The intermediate term now is the harmonic mean $p_z(k) = \frac{2p_z(k-1)p_z(k+1)}{p_z(k-1)+p_z(k+1)}$ of the values of $p_z(k-1)$ and $p_z(k+1)$. Zipf's distribution is also called the discrete Pareto distribution which is used in many social and scientific phenomena.

In one study where the population of cities was considered, a variant of the Zipf's law provided the best fit with the distribution of $p(k) \propto k^{-1.07}$. Zipf saw the principle of least effort at the basis of his eponymous law.

Let δ be the characteristic path length, which is the average shortest path length of all node pairs in the network. Mathematically, whereas for a random network δ has a logarithmic relationship with the total number of nodes N: $\delta \propto \ln N$, for a scale-free network with 2 < λ < 3 it appears to be ln ln $N$ (Cohen and Havlin, 2003).

One approach to understanding long-tailed behavior is through the underlying counting phenomenon. A counting process may be considered to be uniformly distributed over the range {1, …, S}. For a very large number of such processes with random values of S, the set of numbers will satisfy Benford's law and the leading digit $d$ ($d \in \{1, ..., b\}$) occurs with probability (Benford, 1938)

$$p_b(d) = \log_b(1+\frac{1}{d}) \qquad (2)$$

When $b = 2$, the probability of the first digit being 1 is trivially equal to 1. Table 1 compares the first digit frequencies with the corresponding Zipf's law frequencies where we have assumed that the highest frequency for the second case is also 0.301 .

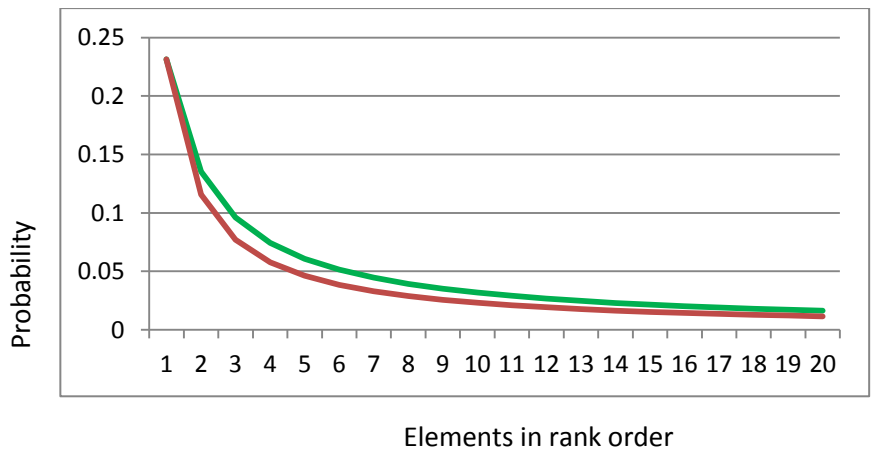

Figure 1. Benford's (20 digits, in red) and Zipf's (green) frequencies

Table 1. First digit frequencies for b=10 and Zipf's law frequencies

|            | 1     | 2     | 3     | 4     | 5     | 6     | 7     | 8     | 9     |
|------------|-------|-------|-------|-------|-------|-------|-------|-------|-------|
| 1st digit  | 0.301 | 0.176 | 0.125 | 0.097 | 0.079 | 0.067 | 0.058 | 0.051 | 0.046 |
| Zipf's freq| 0.301 | 0.155 | 0.103 | 0.075 | 0.060 | 0.051 | 0.043 | 0.038 | 0.034 |

Figure 1 shows how both these distributions are characterized by long tails. It would be correct to assume that if empirical data follows Benford's Law, it is generated by a mixture of independent uniform processes (Janvresse and de la Rue, 2004).

**3. Self-similarity versus combinatorial randomness**
The mathematical approach to randomness is one of combinatorics, whereas biological and natural systems are characterized by self-similarity. The basis of self-similarity appears to be the evolutionary process of recursion across levels and it operates both for internal and external cognitions (Dixon et al., 2014). This self-similarity provides a certain order in randomness that is of value in biological structure, social relationships, as well as evolving physical systems and it can also be useful in checking engineered systems (Nigrini, 2012).

It is of course obvious that an evolving system will not be always perfectly self-similar. Thus the definition of self-similarity must come with a permissible range of values that is appropriate for the system to have the capacity to replicate. Evolution is transformation from one state to another and this requires a potential that drives the transformation.

The dynamic at the basis of evolutionary change in biological networks may be seen in the oscillation between two different modes which defines a kind of a ping-pong model. Such systems appear to have informational and cognitive drivers as in the theory that social size in primates is determined by cognitive capacity. Dunbar, who advanced this theory, argues that the size of the brain is correlated with the complexity of function and he develops an equation, which works for most primates, that relates the neocortex ratio of a particular species - the size of the neocortex relative to the size of the brain – to the largest size of the social group (Dunbar, 1992). For humans, the maximum group size appears to ne 147.8, or about 150, and this represents Dunbar's estimate of the maximum number of people who can be part of a close social relationship.

Support for this idea come from the community of Hutterites, followers of the sixteenth century Jakob Hutter of Austria, who are pacifists and believe in community property and live in a shared community called colony. Several thousand Hutterites relocated to North America in the late 19th century and their colonies are mostly rural. A colony consists of about 10 to 20 families, with a population of around 60 to 150. When the colony's population approaches the upper figure, a daughter colony is established. But for those living in a technology society where the individual's cognitive capacities have been enhanced by a variety of tools one would expect that Dunbar's number would not strictly apply.

Information is also seen as a driver in the evolution of quantum systems (Kak, 2007; Licata and Fiscaletti, 2014; Kak, 2015). Indeed, according to the orthodox interpretation of quantum theory, the very process of observation has an influence on the state of the system.

## 4. Single-layered self-similar system: $p_2$ distribution

Consider a single-layered self-similar system where all members of the network are at the same hierarchical level. In such a system we define self-similarity in terms of the relationship amongst members in adjacent ranks. If the aggregate connectivity is S and the node with the largest connectivity has A links, then the connectivity of the next one is $A/a$, where $a > 1$, with the following one with $A/a^2$, and so on. For each number, the consecutive ratios satisfy the geometric mean $x_n = \sqrt{x_{n-1} x_{n+1}}$.

In other words, the probability distribution with respect to the rank order of connectivity is:

$$p_2(k) = ca^{-k}, \quad k = m, m+1, ..., K \tag{3}$$

This may be contrasted with the power-law distribution (1) where the probabilities are in an inverse rank order. As $k$ becomes large, both this geometric series and the power law distributions become long-tailed, although (3) falls faster. The un-normalized distributions (1) and (3) for values of $\lambda=2$ and $a=2$ with $m=1$ are:

$p_1(k)$: 1, 1/4, 1/9, 1/16, 1/25, 1/36, …

$p_2(k)$: 1, 1/2, 1/4, 1/8, 1/16, 1/32, …

Since the number of nodes is N, we can write for total connectivity generated by $p_2$:

$$S = A(1 + \frac{1}{a} + \frac{1}{a^2} + ... + \frac{1}{a^{N-1}}) \tag{4}$$

This is solved to give:

$$S = \frac{A(a - a^{1-N})}{a-1} > \frac{Aa}{a-1} \tag{5}$$

The value of S will be larger than the right hand side of (5) depending on the length of tail since the minimum number of connections to any node in a connected graph must be 1 and also the constraint that $N > A$ since the first ranked node has A connections.

Since $\frac{dS}{da} = -1/(a-1)^2$, the rate of change of the aggregate number of connections depends on its variance from unity.

A small example of such a network for $A = 10$ and $a \approx 1.4$ is given in Figure 2 below for which the actual number of connections is 40 whereas equation (4) corresponds to the value 35.

In Figure 2, node 1 has 10 connections (for it is connected to all other nodes), node 2 has 7 connections, node 3 has 5 connections, and so on. These are nominal numbers of connections. In a real-world distribution the actual connections will be random with these representing means for a large set of networks.

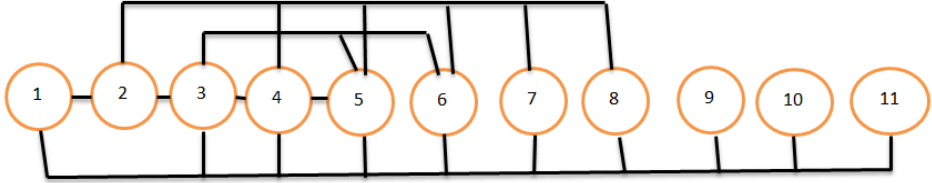

Figure 2. An example of geometric series connectivity with $a \approx 1.4$, A=10

A comparison of this geometric series power law (3) with the Zipf's law is given in Figure 3.

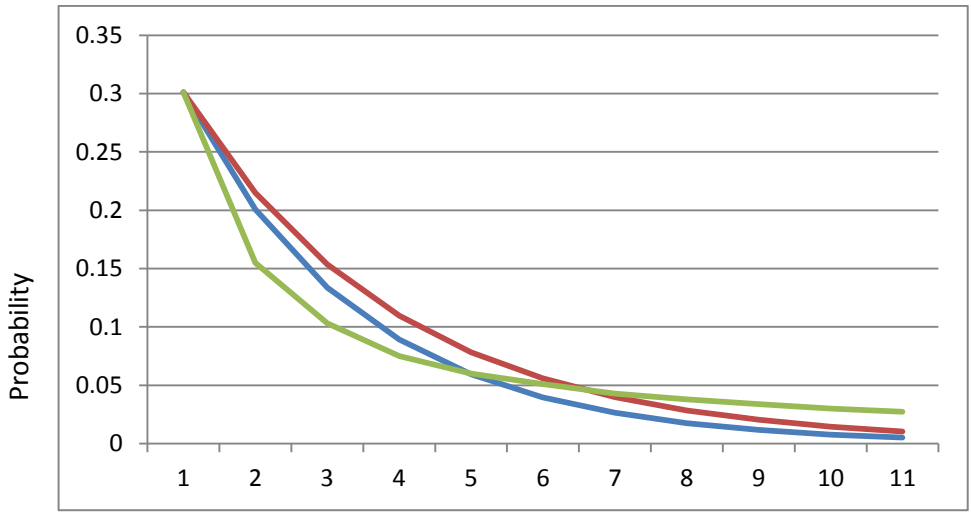

Elements in rank order. Geometric: red: $a$=1.5, blue: $a$=1.4. Zipf: green

Figure 3. Geometric and Zipf distribution frequencies

Long-tailed distributions are characterized by what has been called the 80-20 rule or the Pareto principle associated with the Pareto distribution:

$$p(x) = \frac{\alpha \, x_m^{\alpha}}{x^{\alpha+1}}, \quad x \geq x_m \tag{6}$$

Given that the aggregate connectivity for the geometric distribution is S, we need to find the number of terms $i$, so that for (4)

$$\sum_{i=0}^{i=0.8N} A \frac{1}{a^i} = 0.8S \approx 0.8 \frac{Aa}{a-1} \tag{7}$$

Figure 4 plots the value of *a* for different number of aggregate nodes ranging from 10 to 55 for 20% of the nodes to contribute 80% of the connectivity. As we see the value of *a* goes down as the number of aggregate nodes increases.

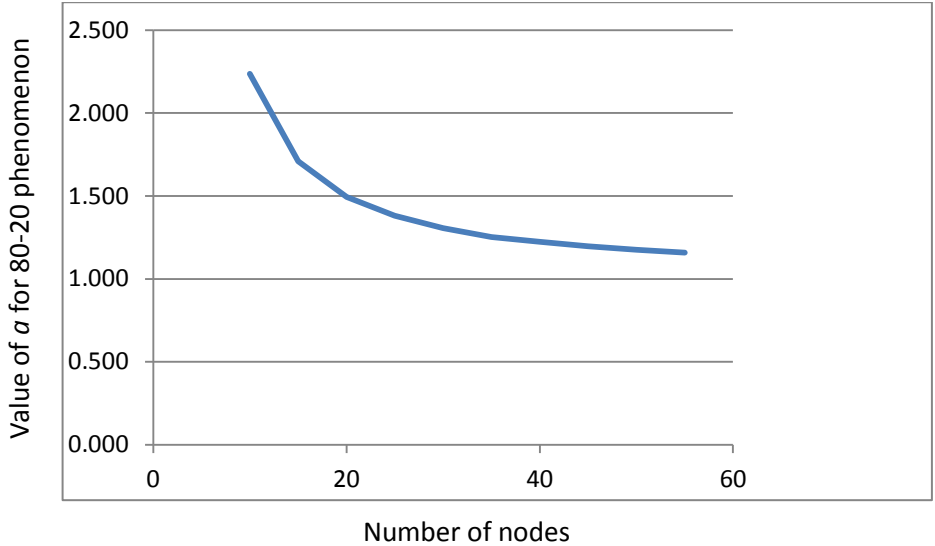

Figure 4. The value of *a* for the 80-20 phenomenon in the power series model

## 5. Hierarchical self-similarity

Now consider self-similarity in a hierarchical sense for the geometric series distribution. The elements in the series

$$1 + \frac{1}{a} + \frac{1}{a^2} + ... + \frac{1}{a^{N-1}} + \frac{1}{a^N} + ... \tag{8}$$

can be put into larger groups such as:

$$(1 + \frac{1}{a} + \frac{1}{a^2}) + \frac{1}{a^3}(1 + \frac{1}{a} + \frac{1}{a^2}) + \frac{1}{a^6}(1 + \frac{1}{a} + \frac{1}{a^2}) + ... \tag{9}$$

Whereas the individual elements have a relationship given by $1/a$, the groups of three have a relationship given by $1/a^3$.

Since for a collection where N is ∞, S= $1/(1-a)$, the relationship between the two is like the one in the mathematical structure of a decimal sequence (Kak and Chatterjee, 1981; Kak, 1985) which is associated with randomness.

## 6. Hierarchical self-similar system: $p_3$ distribution

Now we consider another distribution where the probability of an element (or a group of elements) in a ranked list equals the probability of the next two elements (or the next group of double size) in the sequence shown below:

$$\sum_{i=2^{k-1}}^{2^{k}-1} p_3(i) = \sum_{i=2^k}^{2^{k+1}-1} p_3(i) \tag{10}$$

Specifically,

$$p_3(1) = p_3(2) + p_3(3) = \sum_{i=4}^{7} p_3(i) = ... \tag{11}$$

If the groups are not defined as above (that is associated with powers of 2), then we will have the result (where g represents a group of elements):

$$p_3(g) > p_3(g+1) + p_3(g+2) \tag{12}$$

For a small list, the relation (12) will be approximately correct for many groups. The ranked elements may be seen as constituting hierarchically arranged layers. The relationship between the first ranked element and those of the elements in the second layer will be replicated across the elements of the second and the third layers.

If we wished for the first equality of (10) or (11) to have the same self-similarity as of (3), we have:

$$1 = \frac{1}{a} + \frac{1}{a^2} \tag{13}$$

This implies $a^2 = a + 1$, whose solution is the golden ratio $\varphi$:

$$\varphi = \frac{1+\sqrt{5}}{2} = 1.6180339... \tag{14}$$

For the next group, the relationship will be:

$$\frac{1}{a} + \frac{1}{a^2} = \frac{1}{a^3} + \frac{1}{a^4} + \frac{1}{a^5} + \frac{1}{a^6} \tag{15}$$

This equation $a^5 + a^4 = a^3 + a^2 + a + 1$ can be reduced to $a^4 = a^2 + 1$ by factoring out a+1 on both sides. This indicates how a renormalization of the scaling factor will have to be performed as we go down the hierarchy.

For a system where there is a ternary division (we will associate it with *b*)

$$p_3(1) = p_3(2) + p_3(3) + p_3(4) = \sum_{i=5}^{13} p_3(i) = ... \tag{16}$$

The equation to solve at the next stage for self-similarity of (3) will be:

$$1 = \frac{1}{b} + \frac{1}{b^2} + \frac{1}{b^3} \qquad (17)$$

Its solution is $b=1.8383..$

## 7. Oscillation between two self-similarity measures

Let us consider two states of the system that are associated with different measures of self-replication ($a$ and $b$) in the proposed geometric series model.

$$S_a = A(1 + \frac{1}{a} + \frac{1}{a^2} + ... + \frac{1}{a^{N-1}}) \qquad (18)$$

$$S_b = B(1 + \frac{1}{b} + \frac{1}{b^2} + ... + \frac{1}{b^{M-1}}) \qquad (19)$$

We propose a dynamics bases on the oscillation between these two modes. This constitutes a method that is different from that of preferential attachment which leads to scale-free networks. For simplicity, we consider the distribution (10) rather than (3).

**Example.** Let $a=2$ and $b=3$. Similar to how for $a=2$ the second layer has two elements, for $b=3$, the second layer will have three elements and each of them will have a probability of 1/3 of the first. Let the first stage be governed by $a$. This will create the initial network with connections as below:

Stage 1.

| Nodes | 1 | 2,3 | 4,5,6,7 | Total connections |
|---|---|---|---|---|
| Connections | 4 | 2 each | 1 each | 12 |

This network is realized by the following graph: (1,2), (1,3), (1,4), (1,5), (2,6), (2,7). Stage 2 will be governed by b and it leads to stage 2 below:

Stage 2. Theoretical

| Nodes | 1 | 2,3,4 | 5 to 13 | Total connections |
|---|---|---|---|---|
| Connections | 9 | 3 each | 1 each | 27 |

Stage 2. Actual

| Nodes | 1 | 2,3,4 | 5 to 13 | Total connections |
|---|---|---|---|---|
| Connections | 9 | 4,4,3 | 1 each | 30 |

The theoretical connections in stage 2 are by probabilities alone and they fail the constraint that the total number of connections should be even. Stage 2, Actual, provides a realizable network that is quite close to the theoretical values. Notice, in the transition node 1 has gained 5 new connections, nodes 2 and 3 have gained two each, node 4 has gained 2, nodes 5 through 7 have remained as before, and new nodes 8 through 13 are each connected with one link.

The subsequent growth of the network, swinging between the two different measures of self-similarity, is shown in the tables below, where it is assumed that sometimes some connections are lost to conform to the theoretical numbers:

Stage 3. Theoretical

| Nodes | 1 | 2,3 | 4,5,6,7 | 8-15 | 16-31 | Total connections |
|---|---|---|---|---|---|---|
| Connections | 16 | 8,8 | 4,4,4,4 | 2 each | 1 each | 80 |

Stage 4. Theoretical

| Nodes | 1 | 2,3,4 | 5-13 | 14-40 | Total connections |
|---|---|---|---|---|---|
| Connections | 27 | 9,9,9 | 3 each | 1 each | 108 |

Note that in Stage 4, nodes 5 through 15 have lost one link each.

Stage 5. Theoretical

| Nodes | 1 | 2,3 | 4,5,6,7 | 8-15 | 16-31 | 32-63 | Total connections |
|---|---|---|---|---|---|---|---|
| Connections | 32 | 16,16 | 8,8,8,8 | 4 each | 2 each | 1 each | 192 |

and so on.

We hypothesize that such a process does characterize real social networks. The change in the mode is determined by technology and policy regimes that evolve with time. Therefore, in the real world, *a* and *b* will be time-dependent and the average scaling factor at the second level will appear to be the golden ratio (if the dominant mode is *a*).

**8. Conclusions**
This paper considers the evolution of a social network through the lens of replicating self-similarity at many levels. Closeness to self-similarity in the interconnections is proposed as a measure of the optimality of the organization. Two power series models are proposed to represent self-similarity and they are compared to the Zipf and Benford distributions. In contrast with the Zipf distribution where the middle term is the harmonic mean of the adjoining terms, our distribution considers the middle term to be the geometric mean. A model for evolution of networks by oscillations between two different self-similarity measures is also discussed and it is shown that the scaling at the second level is according to the golden ratio.